\documentclass[prb,twocolumn,showpacs,floats,eqsecnum,amsmath,amssymb]{revtex4}
\usepackage{epsf}
\usepackage{times}
\usepackage[dvips]{graphicx}
\begin{document}

\title{Nonsinusoidal current-phase relations and the $\mathbf{0}-\mbox{\boldmath$\pi$}$ transition in diffusive
  ferromagnetic Josephson junctions}

\author{G. Mohammadkhani and M. Zareyan}

 \affiliation{Institute for Advanced Studies in Basic Sciences,
 45195-1159, Zanjan, Iran}

\date{\today}
\begin{abstract}
We study the effect of the interfacial transparency on the
Josephson current in a diffusive ferromagnetic contact between two
superconductors. In contrast to the cases of the fully transparent
and the low-transparency interfaces, the current-phase relation is
shown to be nonsinusoidal for a finite transparency. It is
demonstrated that even for the nearly fully transparent interfaces
the small corrections due to weak interfacial disorders contribute
a small second-harmonic component in the current-phase relation.
For a certain thicknesses of the ferromagnetic contact and the
exchange field this can lead to a tiny minimum supercurrent at the
crossover between $0$ and $\pi$ states of the junction. Our theory
has a satisfactory agreement with the recent experiments in which
a finite supercurrent was observed at the transition temperature.
We further explain the possibility for observation of a large
residual supercurrent if the interfaces have an intermediate
transparency.
\end{abstract}

 \maketitle

\section{Introduction}

Ferromagnet-superconductor hybrid structures exhibit novel and
interesting phenomena which have been studied extensively in the
recent years (for a review see Refs. \onlinecite{BuzdinRev05} and
\onlinecite{Golubov04}). These systems provide the possibility for
a controlled study of coexistence and competition of the
ferromagnetism and the superconductivity. One of the most
interesting effects is the possibility of forming the so-called
$\pi$ Josephson junction in
superconductor-ferromagnet-superconductor (SFS)
structures\cite{Bulaevskii,Buzdin1,Buzdin2,Buzdin3,Zareyan,
Bergeret,Ryazanov01,Chtchelkatchev1,Radovich03,Chtchelkatchev2}.
In a $\pi$ junction the ground-state phase difference between two
coupled superconductors is $\pi$ instead of $0$ as in the ordinary
$0$ junctions \cite{BuzdinRev05}. The existence of the $\pi$
junction in layered SFS systems, which was first predicted by
Bulaevskii {\it et al.}, \cite{Bulaevskii} and Buzdin {\it et al.}
\cite{Buzdin1} to occur for certain thicknesses and exchange field
energies of the F layer, has been confirmed in
experiment\cite{Ryazanov01}. Ryazanov {\it et al.}
\cite{Ryazanov01} observed the formation of the $\pi$ junction in
diffusive SFS junctions by measuring nonmonotonic variations of
the Josephson critical current on changing the temperature. At the
transition from $0$ to $\pi$ coupling the critical current
undergoes a sign change and thus passes through a zero-current
minimum. This appears as a cusp in the temperature dependence of
the absolute values of the critical current which is measured in
the experiment. Later the nonmonotonic temperature dependence of
the critical current in diffusive SFS contacts was observed in
other experiments
$\cite{Ryazanov02,Guichard03,Kontos02,Sellier03}$ and was
attributed to the $ 0-\pi$ transition induced by the ferromagnetic
exchange field. The $ 0-\pi$ transition have been studied
theoretically by several authors in clean
\cite{Chtchelkatchev1,Bergeret,Radovich03,Chtchelkatchev2} and
diffusive \cite{Buzdin2,Bergeret,Buzdin03,Chtchelkatchev2} F
Josephson contacts for different conditions and strength of
disorder at the FS interfaces. The theory turned out to be in good
agreement with experiment\cite{Belzig05}.
\par
Very recently a new experiment revealed another characteristic of
the $0-\pi$ transition in SFS junctions, which was not detected
before. While in the early experiments
$\cite{Ryazanov01,Ryazanov02,Kontos02,Guichard03,Sellier03}$ the
critical current at the crossover temperature was found to vanish,
Sellier {\it et al.} \cite{Sellier04} reported the existence of a
finite small supercurrent at the transition temperature $
T_{0\pi}$ in diffusive SFS junctions for a certain thickness of
the F layer. A finite critical current at the crossover was
predicted theoretically \cite{Chtchelkatchev1,Radovich03} in
ballistic F contacts, and was attributed to the appearance of the
second harmonic in the Josephson current-phase relation. In Ref.
\onlinecite{Sellier04} further analysis of the residual
supercurrent by applying high-frequency excitations showed
half-integer Shapiro steps, confirming the appearance of the
second harmonic in the Josephson current-phase relation. This
observation was confirmed by other experiments in Ref.
\onlinecite{Ryazanov05}, although it was argued that the
appearance of the half-integer Shapiro steps in the
current-voltage characteristic of the SFS contacts may also
originate from a non uniform thickness of the F layer.
\par
For diffusive SFS contacts the theoretical studies mostly
concentrated on either the perfectly transparent FS interface or
the low-transparency case. For these two limiting cases and in the
weak-proximity-effect regime using the linearized Usadel
equations\cite{Usadel} implemented with the Kupriyanov and
Lukichev\cite{Kupriyanov} (KL) boundary conditions at the FS
interfaces, the current-phase relation (CPR) is found to be always
sinusoidal. Melin \cite{Melin04} was  the first who proposed a
theoretical description of the $\sin(2\varphi)$ term of the CPR in
SFS junctions in which a strong decoherence in the magnetic alloy
can explain the magnitude of the residual supercurrent at the
$0-\pi$ transition. Latter Houzet {\it et al.} \cite{Houzet05} and
Buzdin \cite{Buzdin05} independently presented an explanation of
the nonsinusoidal CPR by considering the effect of magnetic
impurities on the Josephson current in diffusive SFS contacts
using the KL boundary condition in the highly transparent limit.
They showed that the isotropic \cite{Houzet05} or the uniaxial
\cite{Buzdin05} magnetic scattering in the ferromagnet can lead to
a second harmonic component in the CPR which is responsible for
the finite supercurrent at $ T_{0\pi}$. The aim of the present
work is to study the effects of the finite FS interface in the
full range on the $0-\pi$ crossover in diffusive SFS contacts. To
our best knowledge there have not been studies of diffusive F
Josephson junctions with the FS interface having an arbitrary
transparency. We show that even a small correction of the
Josephson current due to weak disorder at the interfaces can
produce a second-harmonic term which leads to a small finite
supercurrent at the $0-\pi$ as was observed in experiment.
\par
In this paper we investigate the effect of the disorder at the FS
interface on the Josephson supercurrent in the diffusive F contact
between two conventional superconductors. We adopt quasiclassical
Green's function method in the diffusive limit implemented by the
general boundary conditions of Nazarov \cite{Nazarov99}, which
allow us to obtain the Josephson current through the contact for
an arbitrary strength of the barrier at the FS interfaces. We
analyze the $0-\pi$ transition and the CPR in terms of the
exchange field and the thickness of the F contact in the full
range of the FS interface transparency. In two limits of high- and
low-transparent interfaces the $0-\pi$ transition occurs on
changing the temperature for different exchange fields and
thicknesses of the F layer. For a transparent interface the
$0-\pi$ transition does not occur for layers of thickness smaller
than the coherence length $\xi=\sqrt{\xi_{0}\ell_ {\rm imp}}$
[here $\xi_{0}=v^{(S)}_{\rm F}/\pi\Delta_0(T=0)$ is the
superconducting coherence length, $v^{(S)}_{\rm F}$ is the Fermi
velocity in the superconductor, $\Delta_0(T=0)$ is the amplitude
of the superconducting order parameter at zero temperature and
$\ell_ {\rm imp}$ is the mean free path in the F layer] even for
very large exchange fields (throughout this paper we use the
system of units in which the Plank and Boltzmann constants
$\hbar=k_B=1$). In contrast to this for a low-transparency
interface the transition takes place for thinner contacts if the
exchange field is strong enough. In these two limiting cases the
current-phase relation is sinusoidal provided that the weak-
proximity approximation holds. This implies a zero supercurrent at
the $0-\pi$ transition.  We show that the corrections to these
results, both in high- and low-transparency cases, produce a
second-harmonic term proportional to $\sin(2\varphi)$. While for
the low-transparency interfaces the second-harmonic term is so
small at it can be neglected, it has some measurable value at the
high-transparency limit which has oscillatory behavior as a
function of the F thickness and the exchange field. For values of
these quantities which lead to a positive value of the
second-harmonic, the critical current at the $0-\pi$ transition
has a small finite value. This finding is consistent with the
experiment\cite{Sellier04}, where evidence for the second harmonic
was found for highly transparent interface samples.
\par
We further show that for interfaces far from being in the high- on
low-transparency regime, the CPR deviates strongly from the
standard sinusoidal form. The second-harmonic term at the $0-\pi$
transition can be as the same order as the first harmonic for an
intermediate value of the FS interface transparency, leading to a
large residual supercurrent at the transition,  provided that the
second-harmonic term is positive for the given exchange field and
thickness of the contact. We analyze the finite supercurrent at
the transition in the full range of the interface's transparency.
\par
In Sec. II, we introduce our model of a diffusive SFS contact.
Assuming the weak-proximity effect regime we linearize the Usadel
equations and the Nazarov boundary conditions at the FS interfaces
with respect to the anomalous component of the quasiclassical
Green's function in the ferromagnetic side of the interface. We
present an expression for the Josephson current which is valid for
an arbitrary barrier strength at the FS interfaces. Section III is
devoted to analyzing the $0-\pi$ transition and the finite
transition current in the full range of the parameters. We obtain
analytical expressions for the second-harmonic corrections in the
two limits of high and low transparencies and also present the
results for intermediate transparency . We also compare our
results with the finding of the recent experiment. Finally we end
with a conclusion in Sec. IV.

\section{The Josephson Current in Diffusive SFS Structures\label{sec2}}

\begin{figure}
\vspace{0.2in} \centerline {\hbox{\epsfxsize=2.5in \epsffile
{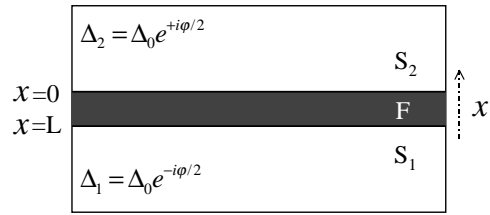}}} \caption{Schematic of the SFS junction. The current
flow between two superconductors (${\rm S_{1,2}}$) through a
ferromagnetic layer (F). The FS interfaces have a finite
transparency.} \label{mzfig1}
\end{figure}

In this section we calculate the Josephson current for a diffusive
SFS junction in which the FS interfaces have an arbitrary
transparency. We consider the SFS junction schematically shown in
Fig. \ref{mzfig1}. It consists of two conventional superconducting
reservoirs ${\rm S_{1}}$ and ${\rm S_{2}}$ which are connected by
a thin ferromagnetic metal (F) layer.  The F layer has a thickness
$L$ and is characterized by a homogeneous spin-splitting exchange
field (energy) $h$ . The Josephson current through the F layer is
determined by the phase difference ${\varphi}$ between the
superconducting order parameters of the two superconductors. For a
diffusive contact the thickness $L$ is much larger than the mean
free path of the impurity scattering, $\ell_{\rm imp}$. Band
mismatches and disorders at the interfaces lead to a
backscattering of the quasiparticles. Thus the ${\rm FS_{1}}$ and
${\rm FS_{2}}$ interfaces have finite transparencies which are
determined by the interface resistances $ R_{b_{0}}$ and $R_{b_{
L}}$, respectively. We consider the limit of large exchange fields
($h\gg T_c$), when the superconducting correlations are strongly
suppressed upon crossing from the superconducting side to the
ferromagnetic side of the FS interfaces. In this case we can
assume a weak superconducting proximity effect and apply a theory
which is linearized with respect to the induced superconducting
order parameter in the F layer.
\par
In the quasiclassical regime $\ell_{\rm imp}$ is much larger than
the Fermi wavelength $\lambda_{\rm F}$ which allows us to use the
quasiclassical Green's function approach. For the diffusive
junction we apply the Usadel equation for the spin-resolved matrix
Green's function $\hat{g}_{\sigma}= \left(\begin{array}{cc}
\begin{array}{c}
g_{\sigma}
\end{array}
&
\begin{array}{c}
f_{\sigma}
\end{array}
\\
\begin{array}{c}
f^{\dagger}_{\sigma}
\end{array}
&
\begin{array}{c}
-g_{\sigma}
\end{array}
\end{array}\right)$,
which in the absence of spin-flip scattering reads
\begin{equation}
[(\omega_{n}+i\sigma h)\tau_{3}+\hat
{\Delta}(\textbf{r}),\hat{g}_{\sigma}]=
D\partial_{\textbf{r}}\hat{g}_{\sigma}\partial_{\textbf{r}}\hat{g}_{\sigma}.
\label{JC:eq1}
\end{equation}
Here $g_{\sigma}$ and $f_{\sigma}$ are, respectively the normal
and the anomalous components of the matrix Green's function for
spin $\sigma~(=\pm1)$ electrons which depend on the Matsubara
frequency $\omega_{n}=\pi T(2n+1)$ and the coordinate
$\textbf{r}$. The diffusion coefficient $D=v^{(F)}_{\rm
F}\ell_{\rm imp}/3$, in which $v^{(F)}_{\rm F}$ is the Fermi
velocity in the F layer and $\tau_{i}$ (i=1,2,3) denote the Pauli
matrices. The matrix ${\hat{\Delta}(\textbf{r})}=
\left(\begin{array}{cc}
\begin{array}{c}
0
\end{array}
&
\begin{array}{c}
\Delta(\textbf{r})
\end{array}
\\
\begin{array}{c}
\Delta^{\ast}(\textbf{r})
\end{array}
&
\begin{array}{c}
0
\end{array}
\end{array}\right)$,
in which $\Delta(\textbf{r})$ is the superconducting pair
potential (order parameter). The matrix Green's function
$\hat{g}_{\sigma}$ satisfies the normalization condition:
\begin{align}
\hat{g}^2_{\sigma}=1&,&g_{\sigma}^{2}+f_{\sigma}f^{\dagger}_{\sigma}=1,
\label{JC:eq2}
\end{align}
where
$f^{\dagger}_{\sigma}(\textbf{r},\omega_{n},h)=f^{\ast}_{\sigma}(\textbf{r},\omega_{n},-h)$.
We have assumed a thin film structure with thickness $L$ much
smaller than the lateral dimensions, so that variations are only
along the $x$ axis in the direction perpendicular to the plane of
the film.
\par
The current density is obtained from $f_{\sigma}$ through the
relation
\begin{equation}
j(\varphi)=\frac{i\pi T}{2e\rho_{F}}\sum_{\sigma=\pm1}
\sum_{\omega_{n}=-\infty}^{\omega_{n}=+\infty}\left(f_{\sigma}\frac{df^{\dag}_{\sigma}}{dx}
-f^{\dag}_{\sigma}\frac{df_{\sigma}}{dx}\right), \label{JC:eq3}
\end{equation}
Here $\rho_{F}$ is the normal-state resistivity of the F layer.
\par
Inside the F layer $\Delta(x)=0$, but $f_{\sigma}$ can still be
nonzero due to the proximity effect. We neglect the suppression of
the order parameter $\Delta$ close to the FS interface inside the
superconductors . Thus $\Delta(T)=\Delta_{0}(T)\exp{(\pm i
\varphi/2)}$ is constant in the superconductors. With this
assumption we solve Eq. (\ref{JC:eq1}) inside the F layer and
impose boundary conditions at the two interfaces which relates the
values of the Green's functions at the superconducting and
ferromagnetic sides of the two interfaces. We apply the general
boundary condition developed by Nazarov \cite{Nazarov99}, which is
valid for an arbitrary value of the barrier height at the
interfaces. For $\hat{g}_{\sigma}$ the boundary conditions are
written as
\begin{eqnarray}
&&\hat{g}_{\sigma} \frac{\partial\hat{g}_{\sigma}}{\partial
x}|_{x=0^{+}(L^{-})}=\frac{R_{F}}{R_{b_{0,L}}L}
\left\langle \hat{M}(\theta)\right\rangle,\label{JC:eq4}\\
\nonumber
\hat{M}(\theta)&=&\frac{{2T_{0,L}}(\theta)[\hat{g}_{\sigma}\mbox{\boldmath$($}0^{-}(L^{-})\mbox{\boldmath$)$}
,\hat{g}_{\sigma}
\mbox{\boldmath$($}0^{+}(L^{+})\mbox{\boldmath$)$}]}{4+T_{0,L}\left(\left\{\hat{g}_{\sigma}\mbox{\boldmath$($}
0^{-}(L^{-})\mbox{\boldmath$)$},
\hat{g}_{\sigma}\mbox{\boldmath$($}0^{+}(L^{+}
)\mbox{\boldmath$)$}\right\}-2\right)},\\
\label{JC:eq5}
\end{eqnarray}
where $R_{F}$ is the normal-state resistance of the F layer. The
transmission eigenvalues of the interfaces at $0,L$ are given by
$T_{0,L}(\theta)=\cos^{2}{\theta}/(
{\cos^{2}\theta+\gamma_{b_{_0,_L}}})$ with the angle $\theta$
showing the direction of the Fermi velocity with respect to the
axis normal to the interfaces,
$\gamma_{b_{_0,L}}=AR_{b_{_0,_L}}/(\xi\rho_{F})$ dimensionless
quantities which measure the strength of the scattering from
disorders at the interfaces $0,L$, respectively, and $A$ the area
of the interfaces\cite{Blonder}. The averaging over directions of
the Fermi velocity is defined via
\begin{equation}
\left\langle
\hat{M}(\theta)\right\rangle=\frac{\int_{-\pi/2}^{\pi/2}d \theta\,
\cos(\theta)\hat{M}(\theta)}{\int_{-\pi/2}^{\pi/2}d \theta\,
\cos(\theta)T(\theta).} \label{JC:eq6}
\end{equation}
$[ ~]$ and $\{ \}$ show the commutator and the anticommutator in
$\hat{M}(\theta)$, respectively. Also
$g_{\sigma}(0^{-})\left[f_{\sigma}(0^{-})\right]$ and
$g_{\sigma}(0^{+})\left[f_{\sigma}(0^{+})\right]$ are normal
(anomalous) Green's functions at the interface $0$ in the
superconductor and the ferromagnet sides, respectively. A similar
notation holds for $g_{\sigma}(L^{\pm})$ and $f_{\sigma}(L^{\pm})$
at the interface $L$. These boundary conditions are valid for
arbitrary values of $ \gamma_{b_{0,L}}$. Equations
(\ref{JC:eq4})-(\ref{JC:eq6}) are reduced to the boundary
condition of KL in the limit  of small transmissions ($
\gamma_{b_{0,L}}\gg1$) and to the condition of the continuity of
the Green's function in the limit of high transparencies
(${\gamma_{ b_{0,L}}\rightarrow0}$).
\par
In a homogeneous bulk s-wave superconductor under the equilibrium
condition, the solutions of the Usadel equations are found to be
$g_{s}=\omega_{n}/{\Omega_{n}}$, $f_{s}=\Delta/{\Omega_{n}}$,
where $\Omega_{n}=\sqrt{\Delta_{0}^{2}+\omega_{n}^{2}}$.
\cite{Aarts} In the diffusive limit, the boundary values of the
Green's functions in the superconductor sides of the interfaces
are approximated by the bulk values:
\begin{eqnarray}
g_{\sigma}\left(0^{-}(L^{+})\right)&=&g_{s}=\frac{\omega_{n}}{\Omega_{n}},
\label{JC:eq7}\\
f_{\sigma}\left(0^{-}(L^{+})\right)&=&f_{s_{1,2}}=\frac{\Delta_{0}}
{\Omega_{n}}e^{\mp i\varphi/2}. \label{JC:eq8}
\end{eqnarray}
In general, the Usadel equation (\ref{JC:eq1}) and the Nazarov
boundary conditions Eqs. (\ref{JC:eq4})-(\ref{JC:eq6}) are
nonlinear with respect to $ g_{\sigma}$ and $ f_{\sigma}$. We
linearize these equations by assuming a weak proximity effect in
the F layer which implies $|g_{\sigma}|\approx1$ and
$|f_{\sigma}|\ll 1$. With this approximation Eqs. (\ref{JC:eq1})
and (\ref{JC:eq5}) are respectively reduced to the following
relations
\begin{equation}
(\omega_{n}+i\sigma h)f_{\sigma}(x,\omega_{n},h)
-\frac{D}{2}\frac{\partial^{2}f_{\sigma}(x,\omega_{n},h)}
{\partial x^{2}}=0, \label{JC:eq9}\\
\end{equation}
\begin{eqnarray}
\nonumber
&&\hat{M}(\theta)=\frac{T_{0,L}(\theta)[\hat{g}_{\sigma}\left(0^{-}(L^{-})\right),
\hat{g}_{\sigma}\left(0^{+}(L^{+})\right)]}{2+(g_{s}
-1)T_{0,L}(\theta)}\\
\nonumber &&\pm\frac{
T^{2}_{0,L}(\theta)f_{\sigma}\left(0^{-}(L^{+})\right)
\left[f_{\sigma}(0^{-}(L^{-}))f^{\dag}_{\sigma}(0^{+}(L^{+}))+c.c.\right]}
{\left[2+(g_{s}-1)T_{0,L}(\theta)\right]^{2}}.\\\label{JC:eq10}
\end{eqnarray}
We note that although Eqs. (\ref{JC:eq9}) and (\ref{JC:eq10}) are
linear with respect to $f_{\sigma}$ in the F layer, they are still
nonlinear in $T_{0,L}(\theta)$. The anomalous Green's function
$f_s $ in the superconducting side of the interface can be of
order unity at low temperatures. Thus the FS interfaces have a
very small width across which $f_{\sigma}$ decreases from the bulk
value $f_s$ to $f_F$ ($f_F$ is the anomalous Green's function  in
the ferromagnet side) which is assumed to be small at large
exchange fields $h\gg T_c$.
\par
The solution of Eq. (\ref{JC:eq9}) inside F layer has the form
\begin{equation}
f_{\sigma}(x,\omega_{n},h)=A_{n\sigma}\cosh(k_{n\sigma}x)
+B_{n\sigma}\sinh(k_{n\sigma}x), \label{JC:eq11}
\end{equation}
where $k_{n\sigma}=\sqrt{2(\omega_{n}+i\sigma h)/D}$. Replacing
the solution Eq. (\ref{JC:eq11}) in Eq. (\ref{JC:eq3}) the
expression for the Josephson current is obtained as:
\begin{eqnarray}
\nonumber I(\varphi)&=&\frac{i\pi TA}{e\rho_{F}}\sum_{\sigma=\pm1}
\sum_{n=-\infty}^{n=+\infty}k_{n\sigma}
Im(A_{n\sigma}B^{\ast}_{n\sigma}).\\ \label{JC:eq12}
\end{eqnarray}
Here $\rho_{F}=1/(2e^{2}N_{0}D)$, where ${N_{0}}$ is the density
of states on the Fermi surface and $Im(~ )$ denotes the imaginary
part. The coefficients $A_{n\sigma}$ and $B_{n\sigma}$ are derived
by applying the boundary conditions (\ref{JC:eq4}) and
(\ref{JC:eq10}) to the solution (\ref{JC:eq11}) and using Eqs.
(\ref{JC:eq7}) and (\ref{JC:eq8}). In this way we have obtained
$A_{n\sigma}$ and $B_{n\sigma}$ and replaced the results in
(\ref{JC:eq12}) to obtain the final expression for the Josephson
current as
\begin{eqnarray}
I(\varphi)=I_{0}\sum_{\sigma=\pm1}\sum_{n=-\infty}^
{n=+\infty}Q_{n\sigma}(\varphi)\sin\varphi, \label{JC:eq13}
\end{eqnarray}
where $I_{0}=2\pi T_{c}/eR_{F}$, $T_{c}$ is the superconducting
critical temperature, and $Q_{n\sigma}(\varphi)$ is given in the
Appendix through Eqs. (\ref{A1})-(\ref{A7}) in which for
simplicity we have assumed $\gamma_{b_{0}}=
\gamma_{b_{L}}=\gamma_{b}$.

\section{DISCUSSIONS AND RESULTS \label{sec3}}

Equation (\ref{JC:eq13}) with $Q_{n\sigma}$ given by Eqs.
(\ref{A1})-(\ref{A7}) expresses the Josephson current through the
diffusive F layer for an arbitrary value of $\gamma_{b}$. In the
limit of $\gamma_{b}\ll 1$ and $\gamma_{b} \gg 1$ the expression
(\ref{JC:eq13}) reduces to the results previously obtained for the
cases of the fully transparent \cite{Buzdin2} and the
low-transparency \cite{Buzdin03} interfaces, respectively. In
these two limits the current-phase relation is sinusoidal. Let us
first find corrections to these results by expanding the Josephson
current Eq. (\ref{JC:eq13}) in terms of powers of $\gamma_{b}$ for
small $\gamma_{b}$ and keeping the terms up to the order of
$\gamma_{b}$. In the opposite limit $\gamma_{b} \gg 1$ we do
expansion in the powers of $1/ \gamma_{b}$ and retain the terms up
to the order of $1/\gamma_{b}^3$. The results in the limits of
$\gamma_{b}\ll 1$ and $\gamma_{b} \gg 1$ have, respectively, the
forms
\begin{eqnarray}
I(\varphi)= I^{(1)}\sin\varphi+I^{(2)}\sin(2\varphi)
\label{D-R:eq1}\\
I(\varphi)= J^{(1)}\sin\varphi+J^{(2)}\sin(2\varphi).
\label{D-R:eq2}
\end{eqnarray}
Here $I^{(1,2)}$ and $J^{(1,2)}$ are, respectively, the first- and
the second-harmonic terms of the CPR in the limits of high and low
transparency and are given in Eqs. (\ref{A8})-(\ref{A13}).
\par
Thus we find that a small deviation from the limits $\gamma_{b}=0$
and the $\gamma_{b} \gg 1$ has two effects on the Josephson
current. First it contributes a small correction to the
first-harmonic current term. Second it produces a small
second-harmonic term. The second-harmonics terms
$I_{n\sigma}^{(2)},J_{n\sigma}^{(2)}$ are oscillatory functions of
$h$ and $L$ and depend on the temperature $T$. In spite of their
smallness they can be important at the $0-\pi$ transition where
the first-harmonic terms vanish. While in the absence of the
second-harmonic terms the critical current undergoes a continuous
sign change at the crossover between $0$ and $\pi$ states, for
nonvanishing and positive values of
$I_{n\sigma}^{(2)},J_{n\sigma}^{(2)}$ the sign change is
discontinuous and the critical current jumps from a positive to a
negative value. The discontinuous transition is manifested as a
nonzero minimum of the absolute values of the critical current at
the $0-\pi$ transition temperature $T_{0\pi}${\cite{Buzdin05}.
\begin{figure}
\vspace{0.2in} \centerline{\hbox{\epsfxsize=3.in \epsffile
{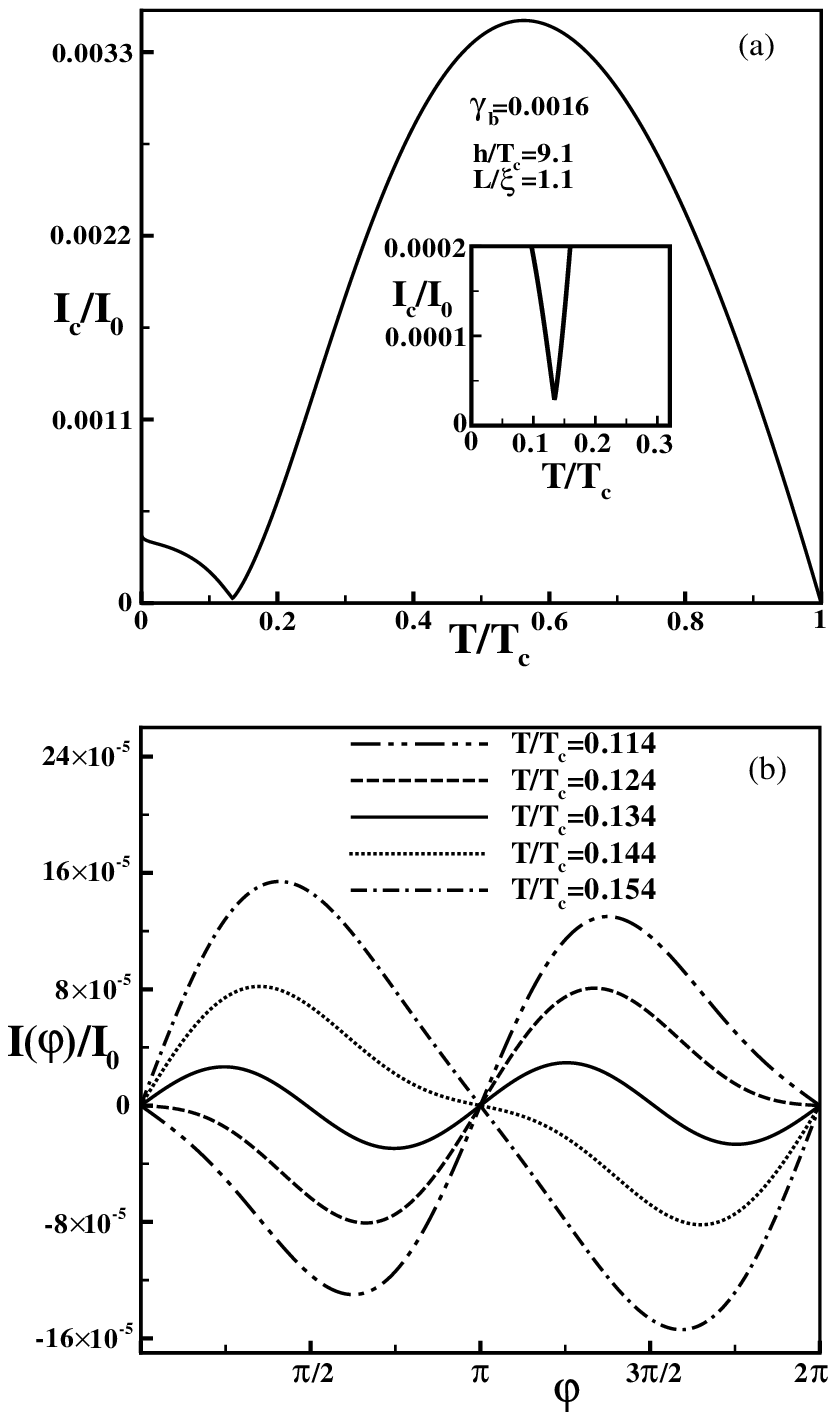}}} \caption{(a) Dependence of the Josephson critical
current on the temperature for $h/T_c={\rm 9.1}$, $L/\xi={\rm
1.1,}$ and $\gamma_{b}={\rm 0.0016}$. Inset shows the temperature
dependence near the transition temperature $T_{0\pi}/T_c={\rm
0.134}$; $I_{\rm c}(T)$ has a nonzero minimum at the transition
temperature. (b) The current-phase relation for different
temperatures around $T_{0\pi}$.} \label{mzfig2}
\end{figure}
\par
In the experimentally relevant limit when ${h\gg T_{c}}$ and ${
L\gtrsim \xi_{F}}$ ($\xi_{F}=\sqrt{D/h}$ is the superconducting
correlation oscillation length in the F layer) we find from Eqs.
(\ref{A8})-(\ref{A10}) that always $ J^{(2)}/J^{(1)}\lesssim {\rm
10^{-5}}$, so the second-harmonic term in the CPR may be neglected
in the low-transparency limit. However for highly transparent
interfaces Eqs. (\ref{A11})-(\ref{A13}) gives us that $I^{(2)}/
I^{(1)}$ can be as large as ${\rm 10^{-3}}$, a ratio that can be
observed in experiment.
\par
Let us now compare our results with the findings of the experiment
of Ref. \onlinecite{Sellier04}. In this experiment the temperature
dependence of the Josephson critical current was studied in the
ferromagnetic alloy layer ${\rm Cu}_{1-x}{\rm Ni}_{x}$ with $x$ of
the order $0.48$, connecting two ${\rm Nb}$ superconducting
electrodes. The normal-resistance measurement of the junctions has
shown that the samples have highly transparent FS interfaces. For
two different thicknesses of the F layer $L_1={\rm 17~nm}$ and
$L_2={\rm 19~nm}$, the $0-\pi$ transition was observed at the
temperatures ${\rm 1.12}$ and ${\rm 5.36~K}$, respectively. For
the sample with $L_2={\rm 19~nm}$ the critical current had a zero
minimum, but they measured a nonzero supercurrent of about ${\rm
4~\mu A}$ for the sample of $L_1={\rm 17~nm}$ at the $0-\pi$
transition temperature. The superconducting critical temperature
of the samples $T_c={\rm 8.7~K}$ and the exchange field was
estimated to be of the order $h={\rm 5-10~meV}$. In Figs.
\ref{mzfig2} and \ref{mzfig3} we present the Josephson critical
current dependence on the temperature and the CPR of the
corresponding two samples using our theoretical results given by
Eqs. (\ref{JC:eq13}) and (\ref{A8})-(\ref{A13}). We fix the
critical temperature of the system, $T_c$, the Fermi velocity $
v^{S}_{\rm F}$ of the superconductors, and the diffusion constant
$D$ of the two F layers with thicknesses $L_{1,2}={\rm 17}$ and
${\rm 19~nm}$, by the values reported in the
experiment\cite{Sellier04}. This leaves two fitting parameters to
be obtained, which are the strength of disorder at the interfaces,
$\gamma_b$, and the exchange field $h$. To take into account the
suppression of the order parameter at the superconducting side of
the FS interfaces, which is more pronounced for strong exchange
fields $h\gg T_c$ and a highly transparent interface, we have
multiplied $\Delta_0(T)$ by a constant factor $\alpha={\rm 0.5}$
in our calculations. This, however, does not change the
qualitative behavior of $I_{\rm c}$ with the temperature, but
allows for a better fitting with the experimental curves. As in
other theoretical studies of the ferromagnetic Josephson junctions
(see e.g., Ref. \onlinecite{Ryazanov01}) the amplitude of our
calculated critical current is substantially ($\thicksim{\rm
10^3}$ times) larger than the measured curves. This can be
attributed to the strong destruction of the Andreev pairs due to
the spin-flip scattering induced by the superparamagnetic behavior
of the Cu-Ni alloy \cite{Sellier03}}, which is not taken into
account in our calculation.
\begin{figure}
\vspace{0.2in} \centerline{\hbox{\epsfxsize=3.in \epsffile
{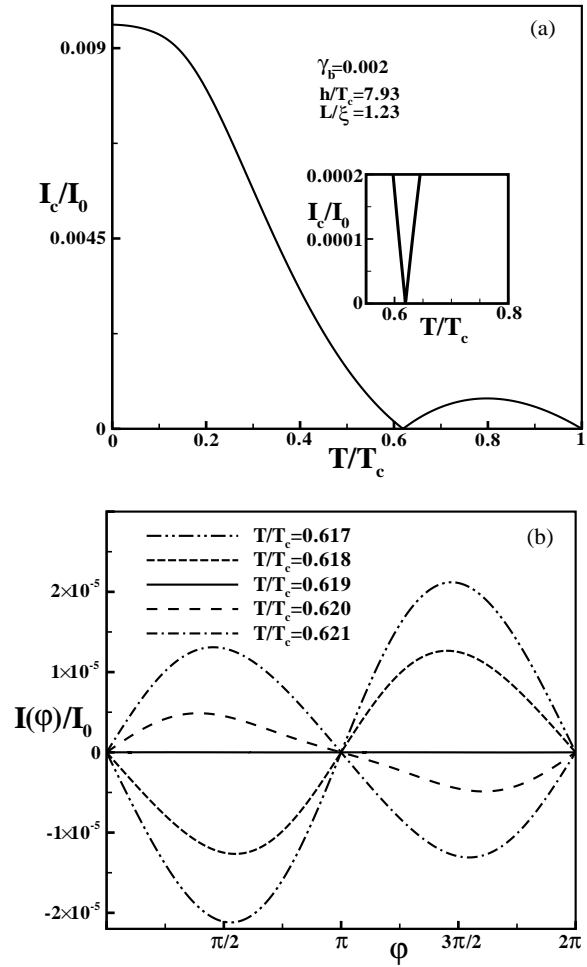}}} \caption{The same as Fig. \ref{mzfig2} but for
$h/T_c={\rm 7.93}$, $L/\xi={\rm 1.23}$, and $\gamma_{b}={\rm
0.002}$.} \label{mzfig3}
\end{figure}
\par
For the thickness ${\rm 17~nm}$ the results are presented in Figs.
\ref{mzfig2}(a) and 2(b). We have taken $D={\rm 5~cm^{2}/s}$ and
$h/T_{c}={\rm 9.1}$. In Fig. \ref{mzfig2}(a) the temperature
dependence of the critical current is presented for one value of
$\gamma_{b}$. There is a good agreement with the experimental
curve at $\gamma_{b}={\rm 0.0016}$. Both the functional dependence
and the transition temperature are reproduced correctly. At the
transition temperature $T_{0\pi}/T_c={\rm 0.134}$ the critical
current has a finite value $I_{\rm c}/I_0\thicksim 3\times {\rm
10^{-5}}$. The finite critical current results from the appearance
of the second-harmonic Josephson current in the CPR. This is shown
in Fig. \ref{mzfig2}(b) in which we present the CPR for different
temperatures around the transition when $\gamma_{b}={\rm 0.0016}$.
Approaching the transition from lower and higher temperatures the
CPR deviates from a sinusoidal form and tends to be a purely
second harmonic proportional to $\sin(2\varphi)$ at $T_{0\pi}$.
\begin{figure}
\vspace{0.2in} \centerline{\hbox{\epsfxsize=3.in \epsffile
{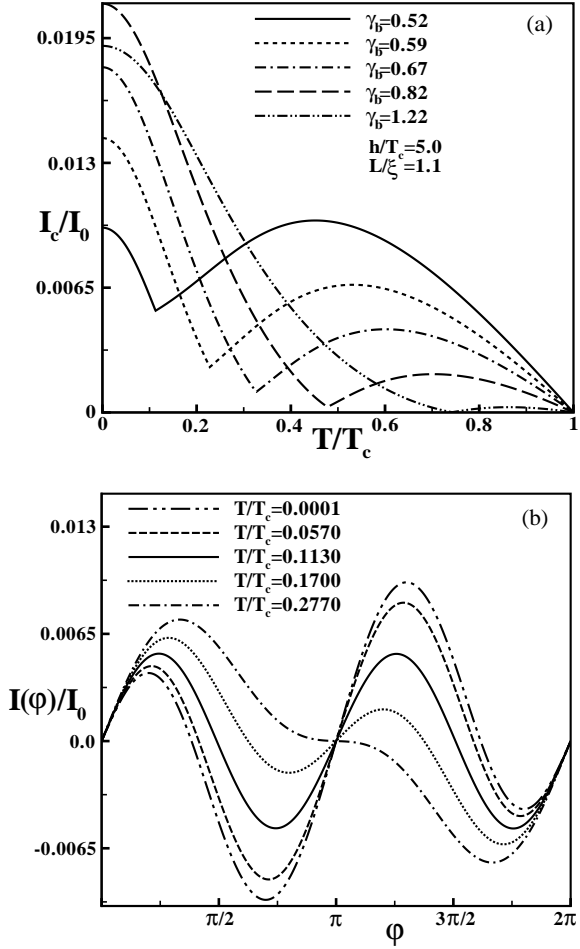}}} \caption{(a) The Josephson critical current versus
$T/T_c$ when $h/T_c={\rm 5}$ and $L/\xi={\rm 1.1}$, for different
$\gamma_{b}$, in the intermediate transparency regime. (b) The
current-phase relation dependence on the temperature for
$\gamma_{b}={\rm 0.52}$.} \label{mzfig4}
\end{figure}
\par
Figure. \ref{mzfig3}(a) shows the temperature dependence of the
critical current for the second thickness ${\rm 19~nm}$. The
diffusion constant and the exchange field have the values $D={\rm
5~cm^{2}/s}$ and $h/T_{c}={\rm 7.93}$. The value of
$\gamma_{b}={\rm 0.002}$ corresponds to the transition observed in
the experiment with $T_{0\pi}/T_c={\rm 0.619}$ and a zero current
minimum. The temperature dependence of the CPR is presented in
Fig. \ref{mzfig3}(b) for $\gamma_{b}={\rm 0.002}$. In contrast to
the $L_1={\rm 17~nm}$ case the CPR remains almost sinusoidal
crossing the transition temperature. At the transition temperature
$I(\varphi)=0$ for all the phases, which represents a continuous
$0-\pi$ transition.
\par
Let us finally analyze the $0-\pi$ transition at intermediate
transparencies. In  Fig. \ref{mzfig4}(a) the temperature
dependence of the critical current $I_{\rm c}(T)$ is presented for
different $\gamma_{b}$ when $h/T_{c}={\rm 5}$ and $L/\xi={\rm
1.1}$. For these values of $h$ and $L$ the $0-\pi$ transition does
not take place for fully transparent interfaces, $\gamma_{b}={\rm
0}$. At a lower transparency $\gamma_{b}={\rm 0.52}$ the
transition occurs at $T_{0\pi}/T_{c}={\rm 0.113}$ with a large
residual minimum supercurrent. On decreasing the transparency,
$T_{0 \pi}$ increases toward the critical temperature $T_c$ and
the transition current decreases and vanishes as $T_{0
\pi}\rightarrow T_c$. Thus we find that the highest values of the
second-harmonic term can be achieved at the intermediate
transparency. This can also be seen from the CPR which is
presented in Fig. \ref{mzfig4}(b) for $\gamma_{b}={\rm 0.52}$ at
different temperatures. There is a strong deviation from a
sinusoidal CPR relation and $I(\varphi)\propto\sin(2\varphi)$
exactly at the transition temperature. We also obtained that for a
fixed thickness by increasing the exchange field the transition
temperature $T_{0\pi}$ shifts toward $T_c$ and the minimum current
amplitude decreases. Similar behavior holds on increasing the
thickness and fixing the exchange field. As a result the $0-\pi$
transition taking place at a higher temperature has a smaller
minimum current. We expect that even for the sample of $L={\rm
19~nm}$ there is an extremely small minimum supercurrent which was
not observable in experiment.

\section{Conclusion \label{sec4}}

In conclusion we have explained that the recently observed finite
supercurrent at the $0-\pi$ transition temperature of diffusive
ferromagnetic Josephson junctions can originate from a weak
quasiparticle backscattering due to small imperfections at the FS
interfaces. Using the quasiclassical Green's functions approach in
the diffusive limit implemented by the general boundary conditions
of Nazarov at the FS interfaces we have obtained an expression of
the Josephson current through the F layer which is valid for any
strength of the interfacial barrier.  At the limiting cases of a
perfectly transparent and a low-transparency interface our results
reduce to the previously obtained expressions of the Josephson
currents with sinusoidal current-phase relations. We have found
that in both limits the corrections to these results contain a
second-harmonic component. While for the experimentally relevant
values of the ferromagnetic layer thickness ($L/\xi_F \gg1$) and
the exchange field ($h/T_c\gg 1$) the second-harmonic correction
is very small and can be neglected in the low-transparency limit,
it can have an observably large positive value for highly
transparent interfaces. We have demonstrated that in the highly
transparent interfaces the positive second-harmonic current
results in a finite supercurrent at the $0-\pi$ transition, which
has been observed in experiment. We have compared our results with
the experimental curves and found a satisfactory agreement. We
also have shown that for an intermediate transparency of the
interfaces the second-harmonic Josephson current can be as large
as the first harmonic, leading to a large residual supercurrent at
the $0-\pi$ crossover in this limit.

\acknowledgments  We acknowledge discussions with D.
Huertas-Hernando, J. P. Morten, Yu. V. Nazarov, and H. Sellier.

\appendix
\section {The expression $Q_{n\sigma}$ in the Josephson current
 for the case $\gamma_{b_{0}}=\gamma_{b_{L}}=\gamma_{b}$}
After deriving the coefficients $A_{n\sigma}$ and $B_{n\sigma}$
and replacing the results in Eqs. (\ref{JC:eq12}) we obtain the
final expression for the Josephson current which is given by Eqs.
(\ref{JC:eq13}), in which $Q_{n\sigma}(\varphi)$ is given by
\begin{eqnarray}
\nonumber
Q_{n\sigma}(\varphi)&=&\frac{32b^{2}LT}{T_{c}}\frac{k_{n\sigma}\beta_{1}\Delta_{0}^{2}}{\beta_{0}}\\
&&\times
\big[\beta_{1}+16c\Delta_{0}^{2}(-\gamma_{b}\xi\cos\varphi+\beta_{2})\big]^{2},
\label{A1}\\
\nonumber
\beta_{0}&=&\big(\beta_{1}^{2}+32c\Delta_{0}^{2}\{\beta_{1}\beta_{2}\\
&&+4c\Delta_{0}^{2}[-k_{n\sigma}\gamma_{b}^{2}\xi^{2}\cos(2\varphi)+\beta_{3}]\}\big)^{2},
\label{A2}\\
\nonumber
\beta_{1}&=&\Omega_{n}^{2}\big[16bg_{s}k_{n\sigma}\gamma_{b}\xi\cosh(k_{n\sigma}L)\\
&&+\left(64b^{2}g_{s}^{2}+k_{n\sigma}^{2}\gamma_{b}^{2}\xi
^{2}\right)\sinh(k_{n\sigma}L)\big],
\label{A3}\\
\nonumber
\beta_{2}&=&k_{n\sigma}\gamma_{b}\xi\cosh(k_{n\sigma}L)+8bg_{s}\sinh(k_{n\sigma}L),\\
\label{A4}\\
\nonumber
\beta_{3}&=&\left(64b^{2}g_{s}^{2}+k_{n\sigma}^{2}\gamma_{b}^{2}\xi
^{2}\right)\cosh(2k_{n\sigma}L)\\
&&+64b^{2}g_{s}^{2}(-1+\frac{k_{n\sigma}\gamma_{b}\xi}{4bg_{s}})\sinh(2k_{n\sigma}L),
\label{A5}\\
b&=& \langle \frac{T(\theta)}{(g_{s}-1)T(\theta)+2}\rangle,
\label{A6}\\
c&=&\langle
\frac{T^{2}(\theta)}{\left((g_{s}-1)T(\theta)+2\right)^{2}}\rangle.
\label{A7}
\end{eqnarray}
\par
In the limits of $\gamma_{b}\ll 1$ and $\gamma_{b} \gg 1$, we can
expand the Josephson current Eq. (\ref{JC:eq13}) in terms of
$\gamma_{b}$. In the limit of $\gamma_{b}\ll 1$, we keep terms up
to the order of $\gamma_{b}$. In the opposite limit of $\gamma_{b}
\gg 1$ we do the expansion in powers of $1/ \gamma_{b}$ and retain
terms up to the order of $1/ \gamma_{b}^{3}$. As a result, the
Josephson current can be expressed in terms of the first and
second harmonics whose coefficients in the limits of
$\gamma_{b}\ll 1$ and $\gamma_{b} \gg 1$ are $I^{(1,2)}$ and
$J^{(1,2)}$, respectively,
\begin{eqnarray}
\frac{I^{(1)}}{I_{0}}&=&\sum_{\sigma=\pm1}\sum_{n=-\infty}^{n=+\infty}
\frac{2LTk_{n\sigma}\Delta_{0}^{2}P_{1}}
{T_{c}\omega_{n}^{2}P_{2}^{2}\sinh(k_{n\sigma}L)},
\label{A8}\\
\nonumber P_{1}&=&1-\gamma_{b}\xi
R_{1}\Delta_{0}^{4}\coth(k_{n\sigma}L),\\
\nonumber
P_{2}&=&\frac{\Delta_{0}^{2}}{(g_{s}+g_{s}^{2})\Omega_{n}^{2}}+2,\\
\nonumber
R_{1}&=&\frac{k_{n\sigma}\left[-8/(\Omega_{n}^{2}\Delta_{0}^{2})+176g_{s}(1+g_{s})+R_{2}\right]}
{4(1+g_{s})^{2}\omega_{n}^{4}P_{2}^{4}},\\
 \nonumber
R_{2}&=&3+\frac{224(1+g_{s})^{2}
\left[1+3\omega_{n}\Omega_{n}(1+g_{s})/(7\Delta_{0}^{2})\right]}
{\omega_{n}^{2}\Delta_{0}^{2}},\\
\label{A9}\\
\frac{I^{(2)}}{I_{0}}&=&\sum_{\sigma=\pm1}\sum_{n=-\infty}^{n=+\infty}
\frac{-2\gamma_{b}\xi
LTk_{n\sigma}^{2}\Delta_{0}^{4}}{T_{c}\omega_{n}^{4}
P_{2}^{3}\sinh^{2}(k_{n\sigma}L)}. \label{A10}
\end{eqnarray}
\begin{eqnarray}
\nonumber
\frac{J^{(1)}}{I_{0}}&=&\sum_{\sigma=\pm1}\sum_{n=-\infty}^{n=+\infty}
\frac{LT\Delta_{0}^{2}R_{3}}
{2T_{c}\gamma_{b}^{2}\xi^{2}\Omega_{n}^{2}k_{n\sigma}\sinh(k_{n\sigma}L)},\\
\label{A11}\\
\nonumber
R_{3}&=&1- \frac{R_{4}\coth(k_{n\sigma}L)}{\gamma_{b}\xi
g_{s}k_{n\sigma}\Omega_{n}^{2}},\\
\nonumber
R_{4}&=&-8g_{s}\Omega_{n}^{2}+\frac{\Delta_{0}^{2}\left[(1+g_{s})-\Delta_{0}^{2}\right]}{(1+g_{s})^{2}}
+\frac{2g_{s}(k_{n\sigma}\xi)^{-2}}{\sinh(k_{n\sigma}L)},\\
\label{A12}\\
\nonumber
\frac{J^{(2)}}{I_{0}}&=&\sum_{\sigma=\pm1}\sum_{n=-\infty}^{n=+\infty}
\frac{-LT\Delta_{0}^{4}}{4(1+g_{s})
T_{c}\gamma_{b}^{3}\xi^{3}\Omega_{n}^{4}k_{n\sigma}^{2}\sinh^{2}(k_{n\sigma}L)}.\\
\label{A13}
\end{eqnarray}

\end{document}